\documentclass[
reprint,
superscriptaddress,
nofootinbib,
amsmath,amssymb,
aps,
pra,
longbibliography
]{revtex4-2}
\usepackage{graphicx}
\usepackage{dcolumn}
\usepackage{bm}
\usepackage{bbm}
\usepackage[caption=false]{subfig}
\usepackage{braket}
\usepackage{amsmath}
\usepackage[T1]{fontenc}
\usepackage[utf8]{inputenc}

\begin{document}
\preprint{APS/123-QED}

\title{Continuous-Variable Multiplexed Quantum Repeater Networks}

\author{Pei-Zhe Li}
\email{p.li@oist.jp}
\affiliation{School of Multidisciplinary Science, Department of Informatics, 
SOKENDAI (the Graduate University for Advanced Studies), 2-1-2 Hitotsubashi, Chiyoda-ku, Tokyo 101-8430, Japan}
\affiliation{National Institute of Informatics, 2-1-2 Hitotsubashi, Chiyoda-ku, Tokyo 101-8430, Japan}
\affiliation{Okinawa Institute of Science and Technology Graduate University, 1919-1 Tancha, Onna-son, Okinawa 904-0495, Japan}

\author{William J. Munro}
\affiliation{Okinawa Institute of Science and Technology Graduate University, 1919-1 Tancha, Onna-son, Okinawa 904-0495, Japan}

\author{Kae Nemoto}
\affiliation{Okinawa Institute of Science and Technology Graduate University, 1919-1 Tancha, Onna-son, Okinawa 904-0495, Japan}
\affiliation{National Institute of Informatics, 2-1-2 Hitotsubashi, Chiyoda-ku, Tokyo 101-8430, Japan}

\author{Nicol\'o Lo Piparo}
\email{nicolo.lopiparo@oist.jp}
\affiliation{Okinawa Institute of Science and Technology Graduate University, 1919-1 Tancha, Onna-son, Okinawa 904-0495, Japan}

\begin{abstract}
Continuous-variable (CV) codes and their application in quantum communication have attracted increasing attention. 
In particular, 
one typical CV codes, cat-codes, has already been experimentally created using trapped atoms in cavities with relatively high fidelities.
However, when these codes are used in a repeater protocol, the secret key rate (SKR) 
that can be extracted between two remote users is extremely low.
Here we propose a quantum repeater protocol based on cat codes with a few quantum memories or graph states as additional resources. This allows us to considerably increase the secret key rate by several orders of magnitude. Our findings provide valuable insights for designing efficient quantum repeater systems, advancing the feasibility and performance of quantum communication over long distances.

\end{abstract}

\maketitle

\section{\label{intro}Introduction}
Quantum communication over long distances is a cornerstone for the advancement of quantum technologies \cite{Duan2001,Jiang2007,Muralidharan2014},
with wide-ranging implications for fields like distributed quantum computing 
\cite{Kimble2008,Cacciapuoti2020}, quantum key distribution \cite{bennett1984proceedings,Ekert1991,Bennett1992}, and secure communication networks \cite{Kimble2008,Chen2009,Wehner2018,Qi2021}. 
However, the transmission of quantum states through optical fibers faces a substantial challenge due to photon loss, severely limiting communication distance and quality of state \cite{Chuang1997,Cochrane1999}. 

Quantum repeaters (QRs) emerge as a solution to this problem, designed to enable the distribution of entangled quantum states over long distances by segmenting the total distance into smaller links 
\cite{Loock2006,Ladd2006,Munro2008,Sangouard2011,VanMeter2014,Munro2015,Azuma2015,Nemoto2016,Azuma2023}.
Now, entangled states generated over such elementary links are stored into quantum memories (QMs)
\cite{Deutsch1996,Briegel1998,Duer1999,Sangouard2011,Jiang2009,Muralidharan2014,Muralidharan2017} and then, through a series of entanglement swapping operations, the entanglement can be ideally extended between two distant users.
However, the implementation of this type of QRs is a daunting task due to the fact that several high-efficiency QMs are required to reach large distances. In fact, low coherence times and small reading/writing efficiencies might deteriorate the stored states, greatly affecting the performance of such systems.
To restore the initial fidelity, a purification mechanism, which requires several QMs \cite{Bennett1996,Deutsch1996,Pan2001,Pan2003,Simon2002,Maruyama2008,Reichle2006,Sheng2008,Wang2011,Wang2015}, must be used, limiting even further the practical realization of the repeater system.
To circumvent the issues with QMs, quantum repeater protocols without the use of QMs have been demonstrated \cite{Fowler2010,Munro2012,Muralidharan2014,Azuma2015,Muralidharan2016,Borregaard2020}.
However, the conventional memoryless protocols typically rely on discrete-variable (DV) error correction codes, which require a big number of single photons \cite{Azuma2015,Borregaard2020}. Since efficient single-photon sources are still far from being realized with the current devices, an alternative approach is given by encoding the information using continuous-variable (CV) codes
\cite{Glancy2004,Loock2006,Dias2017,Furrer2018,Seshadreesan2020}. 
In particular, rotation-symmetric bosonic codes (RSBCs)—a class of CV codes that remain invariant under specific phase-space rotations \cite{Grimsmo2020}—show great potential for their use in quantum repeater protocols due to their high resilience to loss errors. 

Recently, a memoryless quantum repeater scheme based on cavity-QED and RSBCs used to transmit the information over the elementary links has been proposed \cite{Li2023a}. The performance of such a QR scheme has been tested using various RSBCs \cite{Li2024}.
The results show that binomial codes outperform all the other RSBCs analyzed in \cite{Li2024}. 
The generation of binomial codes has been realized experimentally with superconducting systems at microwave frequency \cite{Hu2019,Ni2023}.
Therefore, one must use frequency converters to enable the transmission of such codes over optical fibers by changing their frequency to the telecom frequency used in standard optical fibers.
However, the current frequency converters have low efficiency as well as a large number of them would be required in a quantum repeater system that connects very distant users making the realization of the quantum repeater very challenging.
Besides, although several schemes have been theoretically proposed to generate binomial states within the cavity-QED regime \cite{LoFranco2010,Miry2014}, their experimental realization is still demanding.

It has been shown that the performance of using cat codes is low \cite{Li2024},
the generation of the cat codes within cavity-QED regime has been realized in experiment \cite{Hacker2019} at optical frequencies.
This result makes such states a promising candidate for the implementation of such a CV repeater protocol as they do not require frequency converters.

In this paper, we investigate the main reason behind the low performance of the repeater protocol analyzed in \cite{Li2024} when cat codes are in use.  We then propose a solution to increase the performance of the quantum repeater by slightly modifying this system adding a few QMs at each repeater station and classical communication between two adjacent links.
Alternatively, instead of  QMs, one can think of using small graph states whose nodes are encoded as cat states.
Here we give a theoretical derivation on how such a graph state can be generated only by using  light-matter interaction based on cavity-QED.
To evaluate the performance of the modified quantum repeater protocols, we estimate the secret key rate (SKR) in a quantum key distribution (QKD) scenario in which two distant users are connected by the quantum repeater protocol embedded with cat states in two different cases. In the first case, QMs are used while in the second case, graph states with RSBCs are in use. We compare these two cases under practical assumptions considering the decoherence in the QMs and the probability of creating the entangled states of the graph states.
Then we show that the performance of this quantum repeater scheme is comparable with an existing third generation quantum repeater based on discrete-variable codes \cite{Borregaard2020}.
Our results show that the performance of this repeater system embedded with cat codes can be considerably enhanced by using devices that do not require very high efficiencies and could be realized with the state-of-the-art technology. 

The paper is divided as following: In Section \ref{mc}, we outline the memoryless quantum repeater scheme proposed in \cite{Li2023a} and show the reason why its performance is very low when cat codes are used. Then we introduce a multiplexed approach with additional channels and extra resources: QMs or graph states. In Section \ref{per}, we calculate the SKR in a QKD scenario of the multi-channel protocol and compare them with the single-channel protocol and a recent 3rd generation quantum repeater protocol. In section \ref{con}, we summarize our results.

\section{\label{mc}The Quantum Repeater Protocol}

\begin{figure}[]
    \center
    \subfloat[]{
        \includegraphics[width=0.47\textwidth]{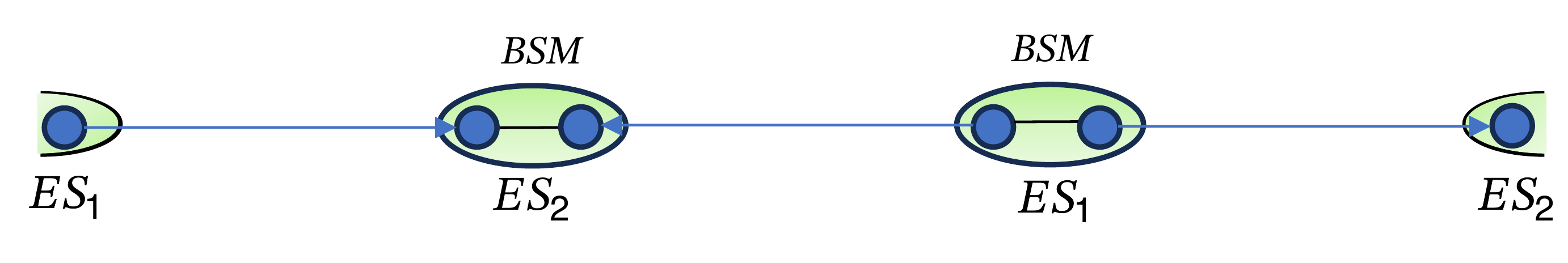}
        \label{figsc}}
    \hfill
    \subfloat[]{
        \includegraphics[width=0.47\textwidth]{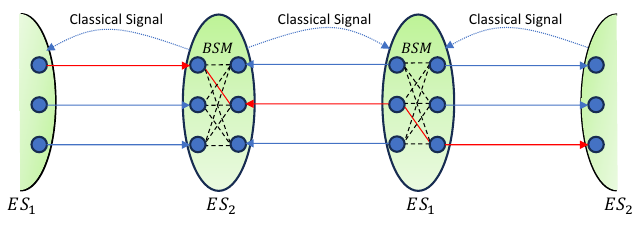}
        \label{figmcm}}
    \hfill
    \subfloat[]{
        \includegraphics[width=0.47\textwidth]{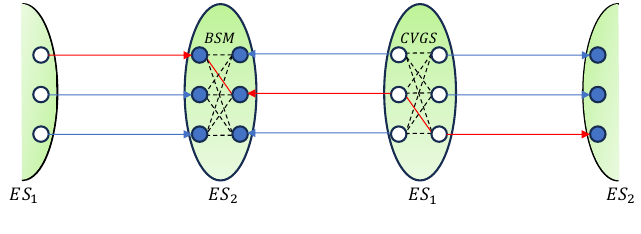}
        \label{figmcml}}
    \caption{\label{figmc} 
    (a) The main building blocks of the memoryless quantum repeater scheme based on cavity-QED and RSBCs using one channel. The blue circles are the trapped atoms/ions in cavities and the blue arrows indicate the propagation directions of the light modes.
    The schematic illustration of the multi-channel protocol embedded with QMs and graph states are shown in (b) and (c), respectively.
    The blue dots refer to the matter qubits in the elementary 
    stations and the white circles are the light modes.
    In (b), the black dotted lines in both $ES_1$s and $ES_2$s are the BSMs that can be locally performed between the matter qubits upon a desired syndrome measurement outcome while the red solid lines are the actual BSMs performed between matter qubits entangled with the light modes for which the corresponding syndrome measurement has a desired outcome.
    The blue dotted arrows in (b) refer to the classical 
    communication occurring between the elementary stations.
    In (c), the black dotted lines and the red solid lines in $ES_2$ refer to the same elements as for (b). In $ES_1$, the white circles are the nodes of the graph states and
    the black dotted lines with the red solid line in $ES_1$ show the connections when the continuous-variable graph state (CVGS) is created in each $ES_1$. Here the black dotted lines are the connections discarded while the red solid line is the one that remains with the desired outcome of syndrome measurements. 
    The solid arrows represent the transmission of the light modes and 
    the red ones indicate the selected ones with preferred syndrome 
    measurement outcomes. 
    }
\end{figure}
In this section, we delineate the reason as to why the quantum repeater protocol analyzed in \cite{Li2024} has low performance when cat codes are in use. 
This will motivate the use of QMs and graph states to enhance its performance.
We then describe into details how these two setups work and how they can be realized.

Let us first describe the main building blocks of the quantum repeater protocol under study, as shown in Figure \ref{figsc}. A channel with a total distance $L_{tot}$ connecting two remote users is divided into smaller segments of elementary distance $L_0$, in which entangled states are created between two adjacent elementary stations, as shown in Figure \ref{figsc}.
Based on their task, we can distinguish two different types of elementary stations, labelled $ES_1$ and $ES_2$ in Figure \ref{figsc}.
At $ES_1$s, 
the entangled light modes encoded with the RSBCs are generated and 
the light modes propagate toward opposite directions to the nearest $ES_2$ stations.
In each $ES_2$, a syndrome measurement is performed to detect the loss errors. 
This syndrome measurement is performed on ancillary qubits entangled with the light modes and ideally enhances the fidelity of the state by detecting the parity of the number of photons lost in the transmission. Once the syndrome measurement has been successfully performed, the light mode is entangled with another matter qubit and is measured in the logic Z-basis.
In both $ES_1$ and $ES_2$s, Bell state measurements (BSMs) are immediately performed between the matter qubits to swap the entanglement and ideally reaching the distance between Alice and Bob.
It is worth noticing that the generation of the entangled light modes, the syndrome measurements and the entanglement creation at $ES_2$s are all based on the interaction of the light modes and matter qubits. 

\begin{figure}[hbt]
    \center
    \includegraphics[width=0.49\textwidth]{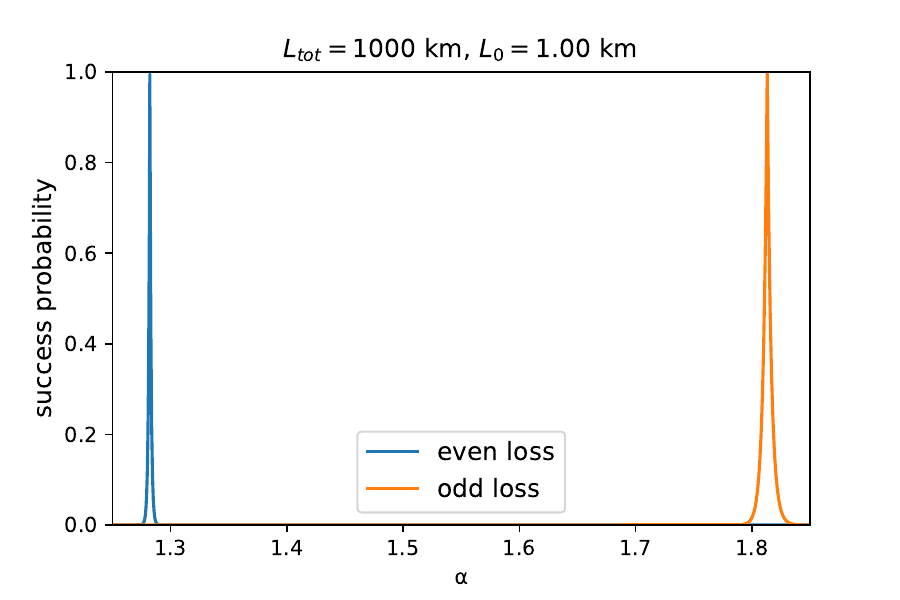}
    \caption{\label{figcpeo} The total success probability for the 1-loss cat codes versus 
    $\alpha$ 
    for different loss events: even photon loss (blue curve) and odd photon loss (orange curve). The total distance $L_{tot}=1000$ km and the elementary distance $L_0=1$ km.
    }
\end{figure}

Due to the fact that the codewords of the cat codes are not orthogonal, the success probability of discriminating such states unambiguously is lower than one.
Besides, this success probability is strongly affected by the average number of photons contained in the state, parametrized by the value of $\alpha$. 
Therefore, the total success probability that Alice and Bob  share an entangled pair, given by the product of  the success probabilities at each elementary station $ES_2$, might considerably  decrease with the choice of the value of $\alpha$. Figure \ref{figcpeo} shows the total success probability versus $\alpha$ when cat codes are in use. 
Here one can see that the total probability of success has two peaks for two different values of $\alpha$, which correspond to a specific syndrome measurement outcome. Specifically, the blue curve is peaked at a certain value of $\alpha$, which corresponds to an even loss event detected by a syndrome measurement whereas the orange curve is peaked at a different value of $\alpha$, which corresponds to an odd loss event.
In general, the average 
number of photons to be used will be a trade-off among all the possible outcomes of the syndrome measurement performed at each $ES_2$ station. This average number of photons severely decreases the total success probability. 

In order to increase the probability of a favorite outcome of a syndrome measurement, one can 
think of adding multiple identical channels connecting $ES_1$ and $ES_2$, in which multiple copies of RSBCs travel. This will increase the probability of having a desired outcome of a syndrome measurement. In this way, one can select the channels in which a specific outcome has occurred. This allows to maximize the success probability by picking a specific numerical value of $\alpha$. 
The channels in which the desired syndrome measurement outcomes occur at $ES_2$s will be communicated to the nearest $ES_1$ station, in which a 
BSM will be performed between the relevant matter qubits. For this setup, QMs are necessary to store the quantum states at the $ES_1$ stations
until the results of the desired syndrome measurement outcome has classically reached the $ES_1$ station. 
However, one can think of substituting the QMs with continuous-variable graph state (CVGS) 
in each $ES_1$ stations. These graph states can be considered as an extension to higher dimensional 
entangled states of the Bell pairs \cite{Briegel2001,Raussendorf2003,Duer2003,Hein2004,VandenNest2004,Aschauer2005} created in the single-channel protocol. 
In the following subsections, we will describe how to create the entangled states with only desired outcome of the syndrome measurements for setups using QMs and 
graph states, respectively.
\subsection{\label{qm}Multi-Channel Protocol Using QMs}
Figure \ref{figmcm} illustrates the setup of the multi-channel quantum repeater protocol with quantum memories.
In $ES_1$, multiple copies of cat-codes are entangled with matter qubits and sent to the nearest $ES_2$ stations.
Unlike the conventional protocol in which these matter qubits are immediately measured, now their states are stored into QMs. 
The light modes will then interact with  ancillary qubits at $ES_2$ on which a syndrome measurement is performed. 
The channels corresponding to a desired syndrome measurement outcome (red lines in Figure \ref{figmcm}) will be communicated to the nearest $ES_1$ stations 
whereas the other light modes will be discarded.
After the classical communication to $ES_1$s, a BSM can be now performed to  the corresponding matter qubits in $ES_1$ (red lines at each $ES_1$ node). Note that QM is unnecessary in $ES_2$s because the syndrome measurement outcomes are local to $ES_2$s and the BSM can be performed immediately after the light modes are measured without storing the atomic states into memories.
In the end, the entangled state shared by Alice and Bob 
is created by swapping those entangled states for which a desired syndrome measurement outcome has occurred. 
This allows to pick a value of $\alpha$ that maximizes the total success probability.

\subsection{\label{cs}Multi-Channel Protocol Using Graph States}
An alternative way of increasing the total probability of success is using graph states at each elementary station, as shown in Figure \ref{figmcml}. Here the nodes of a single graph state represent the logical states of cat codes, which are created in $ES_1$ (white dots in Figure \ref{figmcml}). Then, these nodes propagate across the elementary channels until they reach $ES_2$. The light modes will then be entangled with ancillary matter qubits on which a syndrome measurement is performed. 
We assume that at least one light mode on both side of the graph gives a desired outcome of the syndrome measurement. A logic Z-basis measurement will then be performed on the light modes corresponding to an undesired outcome of a syndrome measurement. This has the effect of removing the corresponding nodes from the graph and creating a bipartite entangled state \cite{Raussendorf2001}.

\begin{figure}[hbt]
    \center
    \includegraphics[width=0.45\textwidth]{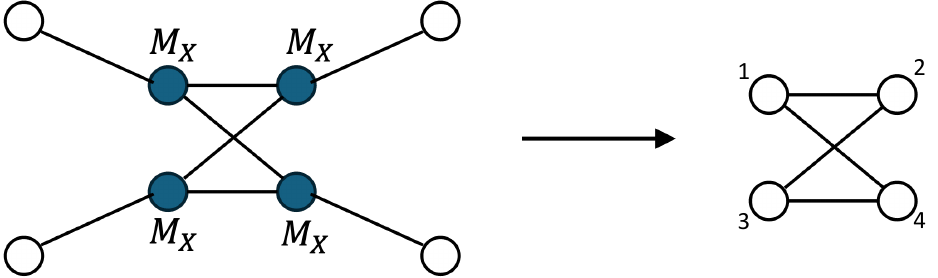}
    \caption{\label{figcs} Creation of graph states. The matter qubits (blue nodes of the graph state on the left) are entangled with encoded light modes (white dots) using the light-matter interaction based on cavity-QED. Then an X-basis measurement performed on these matter qubits will project the light-modes into the graph state illustrated in the right part of the figure.
    }
\end{figure}

The CVGS can be created with the same light-matter interaction based on cavity-QED \cite{Hacker2019} and used to perform all the main building blocks of the protocol. 
At first, a matter-qubit graph state made by trapped atoms/ions in cavity QED is created using locally multiple CZ gates between the matter qubits.
Then each matter qubits in the graph will be entangled with a light mode encoded with RSBCs by using the light-matter interaction based on cavity-QED \cite{Li2023a}.
A hybrid light-matter graph is then created (see Figure \ref{figcs}). Finally, all the matter qubits will be measured on the X basis to remove them obtaining, thus, a graph state of photonic modes in which each node is a RSBCs and the edges are entangling gate between those nodes.
More details about the mathematical derivations can be found in Appendix \ref{cvrs}.

\subsection{\label{cl}Secret Key Rate Analysis}
In this subsection, we derive the mathematical expression of the secret key rate for the multi-channel quantum repeater protocol in which quantum memories and graph states are used, respectively.
The secret key rate of the BBM92 protocol \cite{Bennett1992} applied to a single channel repeater scheme is given by
\begin{eqnarray}
    \label{eq2.1}
    R_{QKD}:=R_{raw}r_{\infty},
\end{eqnarray}
where $R_{raw}$ is the raw key rate and $r_{\infty}$ is the secret fraction \cite{Scarani2009}. 
Now
the raw key rate can be written as
\begin{eqnarray}
    \label{eq2.2}
    R_{raw}=P_{tot}/t_r,
\end{eqnarray}
where $P_{tot}$ is the total success probability and $t_r$ is the repetition time of the whole system, respectively.
In the multi-channel protocol, $P_{tot}$ is given by
\begin{eqnarray}
    \label{eq2.3}
    P_{tot}=P_{tdsm}P_{tz},
\end{eqnarray}
where $P_{tdsm}$ is the probability to obtain desired syndrome measurement outcomes and $P_{tz}$ is the probability that 
all the logical Z measurements of the cat codes are successful in the system. 
For simplicity, we assume that at least one syndrome measurement will give the desired outcome. Therefore the $P_{tdsm}$ is given by
\begin{eqnarray}
    \label{eqc2.1}
    P_{tdsm}=[1-(1-P_{dsm})^m]^n,
\end{eqnarray}
where $P_{dsm}$ is the probability to obtain the desired syndrome measurement outcome in a single channel of an elementary link while $m$ is the number of channels used in each elementary link and $n=L_{tot}/L_0$ is the number of elementary links of the system.

The logic z-basis measurement is performed by an unambiguous state discrimination (USD) \cite{Dieks1988,Peres1988}. This type of quantum measurement might result in a failure event, in which the logic states cannot be discriminated.
The optimal success probability of USD is upper bounded by \cite{Dieks1988,Peres1988}
\begin{eqnarray}
    \label{eqc2.2}
    P_{USD}=1-|\braket{\bar{0}|\bar{1}}|,
\end{eqnarray}
where $\ket{\bar{0}}$ and $\ket{\bar{1}}$ are the codewords of the cat codes. 
Further we assume that the success probability will depend on a measurement error probability $1-p_m$  as
\begin{eqnarray}
    \label{eqc2.3}
    P_{tz}=(P_{USD}p_m)^{k_m},
\end{eqnarray}
where $k_m$ is the number of the logical Z measurements performed in the system. We have $k_m=n$ for the protocol using QMs and $k_m=nm$ for the protocol using graph states.

While the repetition rate in the single channel protocol is given by the light-matter interaction time $t_0$, for the multi-channel protocol with QMs we need to include the communication time as well. For such a protocol $t_r$ is given by
\begin{eqnarray}
    \label{eq2.4}
    t_{r,2}=max\{t_0,2L_0/c\},
\end{eqnarray}
where 
$c$ is the speed of light. On the other hand, in the multi-channel protocol with graph states, we have
\begin{eqnarray}
    \label{eq2.5}
    t_{r,3}=t_0,
\end{eqnarray}
because there is no need to wait for the classical signal.

Next the secret fraction can be written as
\begin{eqnarray}
    \label{eq2.6}
    r_{\infty}=1-h(e_z)-h(e_x),
\end{eqnarray}
where $h(p)$ is the binary entropy $h(p)=-p\log_2{p}-(1-p)\log_2{(1-p)}$ with $(e_x, e_z)$ being the quantum bit error rates (QBERs) \cite{Abruzzo2013,Kirby2016}. As for the cat codes in the multi-channel protocols, it can be shown that $e_z=0$ and $e_x=1-F_{tot}$ if only loss errors are considered, where $F_{tot}$ is the final fidelity of the entangled atomic state shared by Alice and Bob with the target state being the corresponding Bell state. 
The analysis of the fidelity when graph states are used can be found in Appendix \ref{acs}.
When QMs are used, we consider the additional errors caused by the finite coherence time of the QMs. The errors are modeled by considering either depolarizing or dephasing channels. Thus, the total fidelity of the multi-channel protocol using QMs, $F_{tot,2}$, will be lower than the total fidelity of the multi-channel protocol using graph states, $F_{tot,3}$. In addition, $e_z=0$ does not always hold, thus, $e_z$ has to be included in the calculation of secret fraction. More details can be found in the Appendix \ref{cvf}.

The SKR of the multi-channel protocol with QMs is
\begin{widetext}
    \begin{eqnarray}
        \label{eq2.7}
        R_{QKD,2}=\frac{[1-(1-P_{dsm})^m]^n(P_{USD}p_m)^n[1-h(e_{z,2})-h(e_{x,2})]}{max\{t_0,2L_0/c\}},
    \end{eqnarray}
where $e_{z,2}$ and $e_{x,2}$ refer to the QBERs for the protocol using QMs. The SKR of the multi-channel protocol with graph state is
    \begin{eqnarray}
        \label{eq2.8}
        R_{QKD,3}=\frac{[1-(1-P_{dsm})^m]^n(P_{USD}p_m)^{mn}[1-h(1-F_{tot,3})]}{t_0}.
    \end{eqnarray}
\end{widetext}
Finally, to take into account the number of channels in use, we evaluate the SKR normalized over the number of channels $m$,
which is given by $R_{nQKD}=R_{QKD}/m$. Then the total SKR per channel use can be written as $\mathcal{R}=t_rR_{nQKD}$.

\section{\label{per}Performance}

\begin{figure}
    \center
    \includegraphics[width=0.49\textwidth]{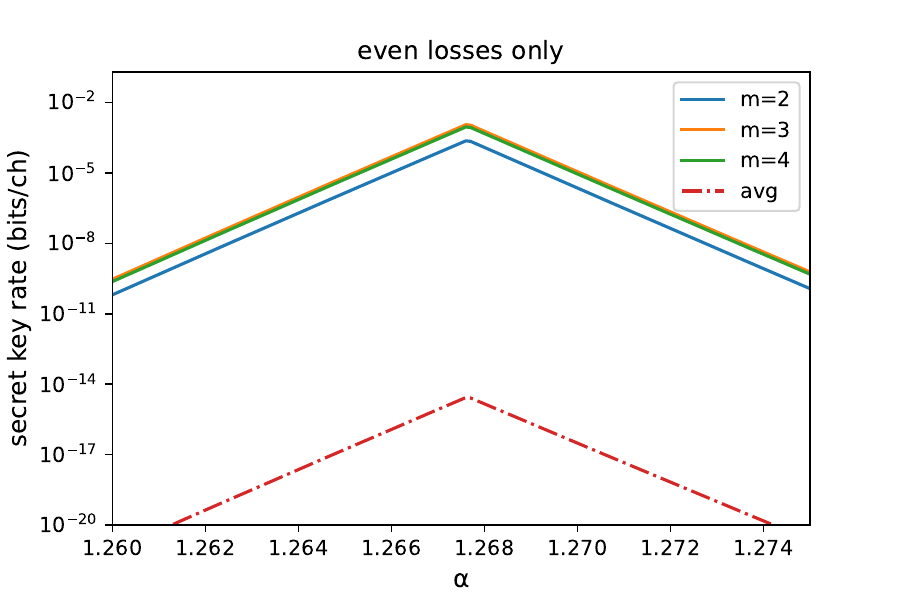}
    \caption{\label{figmcvsc} Comparison of the SKR of the multi-channel protocol for different values of $m$ and the single-channel protocol with the 1-loss cat codes versus $\alpha$. In the multi-channel protocol, the desired syndrome measurement outcome is chosen to correspond to an even number of photon being lost. Here $m$ is the number of channels used in the multi-channel protocol and "avg" refers to the average SKR of single-channel protocol, which is averaged over all possible syndrome measurement outcomes at each elementary link.}
\end{figure}

\begin{figure}
    \center
    \includegraphics[width=0.49\textwidth]{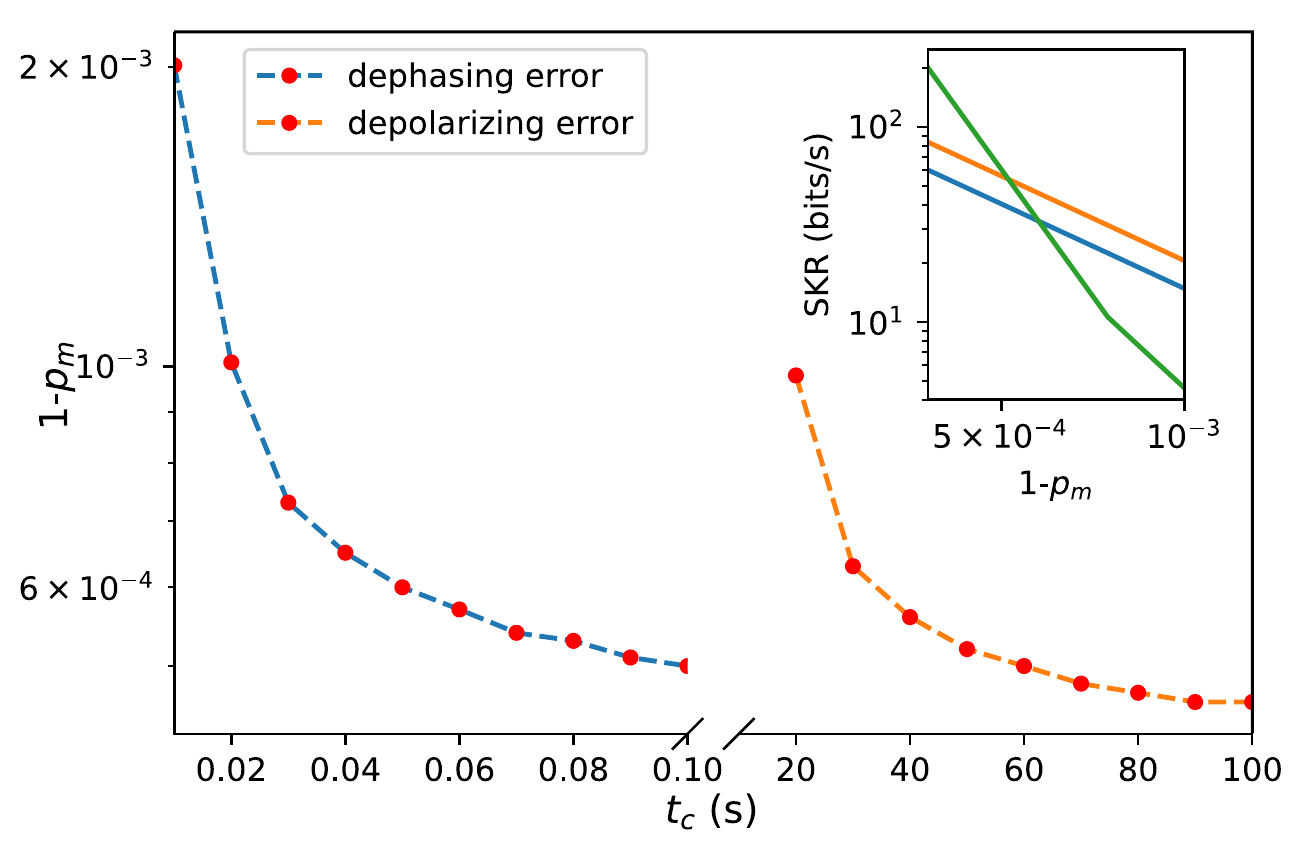}
    \caption{\label{figmcp} The 
    measurement error probability $1-p_m$ 
    versus the coherence time of the QMs corresponding to case in which the SKR of the setup with QMs is equal to the one with graph states (red dots). Both 
    dephasing and depolarizing errors are considered for the QMs. 
    The total distance $L_{tot}=1000$ km and the elementary
    distance $L_0=0.5$ km. The inset shows the optimized SKR over the value of mean photon number and the number of channels for the multi-channel protocol 
    using two different approaches (using QMs or graph states) with optical states encoded in 1-loss cat code
    versus $1-p_m$. 
    The green solid curve shows the SKR of the 
    setup with graph states. The blue curve shows the SKR for the setup with QMs affected by dephasing errors
    at a coherence time $t_c=0.05$ s while the orange solid curve shows the SKR of the setup with QMs affected by depolarizing error 
    at a coherence time $t_c=50$ s.}
\end{figure}

In this section, we compare the performance of the multi-channel repeater protocol with QMs and with graph states under practical assumptions. To this end we numerically calculate 
of both systems using Eqation \ref{eq2.7} and \ref{eq2.8}, respectively.
At first, we show that the performance of the multi-channel protocol in the ideal case is significantly improved compared to the single-channel protocol.
The performance of the
two different setups of the multi-channel protocol is then compared including the effects of the coherence time of the QMs, $t_c$, and the measurement error probability, $1-p_m$.

Figure \ref{figmcvsc} shows the SKR of the multi-channel protocol and the single-channel protocol when the 1-loss cat codes are used versus $\alpha$ for a different number of channels. One can see that the SKR is significantly increased when the multi-channel protocol is considered. For instance, at $\alpha\approx1.268$, the SKR of single-channel protocol is near $10^{-14}$ bits/ch while the SKR of multi-channel protocol can reach $10^{-3}$ bits/ch when $m=3$. These results show that the SKR can be increased more than 10 orders of magnitude and only a few additional channels are needed for this improvement.

We now compare the two multi-channel protocols when their major source of errors (dephasing or depolarizing in the QMs and measurement errors in the graph states) are considered. 
Figure \ref{figmcp} shows the SKR versus the measurement error of the multi-channel protocol with graph states (green curve) and with QMs in which decoherence is modeled by a dephasing (depolarizing) channel by the blue (orange) curve, respectively. The crossing points in the inset correspond to the cases in which the SKR of the multi-channel protocol with QMs is equal to the one with graph states. In the main figure, these crossing points are represented by the red dots. The dashed lines connecting the dots have been shown to refer to the two types of decoherence occurring in the QMs, blue lines for the dephasing process and orange line for a depolarizing process, respectively. 
The SKR of the setup with QMs (setup 1) is higher than the setup with graph states (setup 2) when the measurement error is above a certain threshold. For instance for a dephasing error model
at $t_c=0.05$ s, the SKR of setup 1 is higher than setup 2
with $1-p_m>6\times10^{-4}$ whereas for $1-p_m<6\times10^{-4}$, as shown in the inset of Figure \ref{figmcp},
setup 2 is more efficient.
We find that for higher coherence time, the required measurement error to achieve the same SKR becomes lower. 
This clearly shows a notable trade-off between measurement error and coherence time in quantum memories. 
\begin{figure}
    \center
    \includegraphics[width=0.49\textwidth]{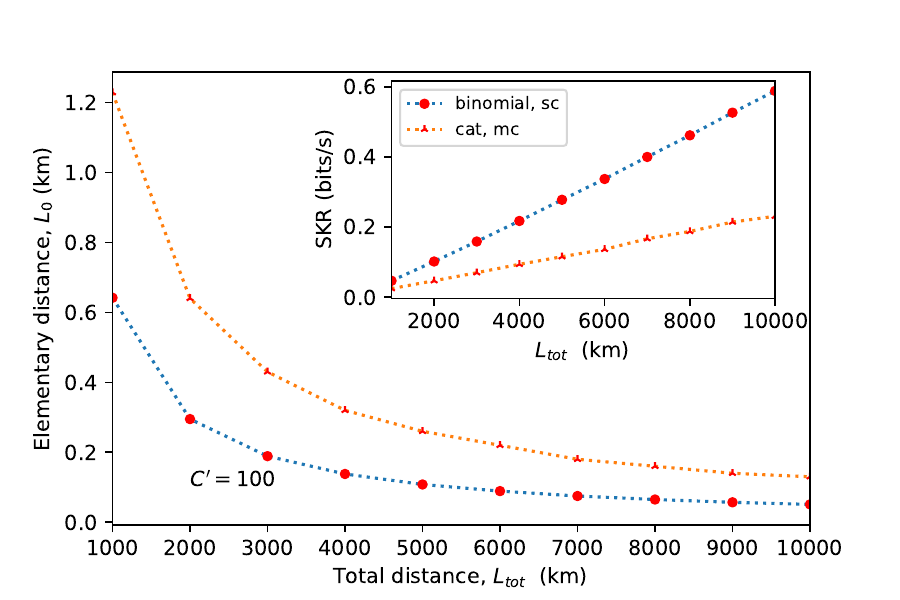}
    \caption{\label{figcpmc} The elementary distance versus the total distance between Alice and Bob required to obtain a cost coefficient $C^\prime=100$ when the single channel protocol with binomial codes is in use (triangular red dots) and when the multi-channel protocol with cat codes is used (round red dots). The inset shows the corresponding SKRs in bits/s for both protocols.}
\end{figure}

Then we compare the multi-channel quantum repeater protocol embedded with QMs with the conventional single-channel protocol in which binomial codes are in use. Here the cost coefficient $C^\prime$ is defined as $C^\prime=C/L_{tot}$, where $C=N_{tot}/R_{QKD}=N_s L_{tot}/R_{QKD} L_0$ \cite{Muralidharan2014}.
$N_{tot}$ is the total number of matter qubits (trapped atoms in cavity-QED regime) included in the protocol while $N_s$ is the number of matter qubits within each elementary link.
For both schemes the logic states used to transmit the information are 1-loss codes that can detect and correct the loss of one photon. We also assume that the quantum states stored into the QMs undergo a dephasing channel with coherence time $t_c=0.5$ s. The measurement error is not considered for both schemes to provide a fair comparison.
\begin{figure}[h]
    \center
    \includegraphics[width=0.49\textwidth]{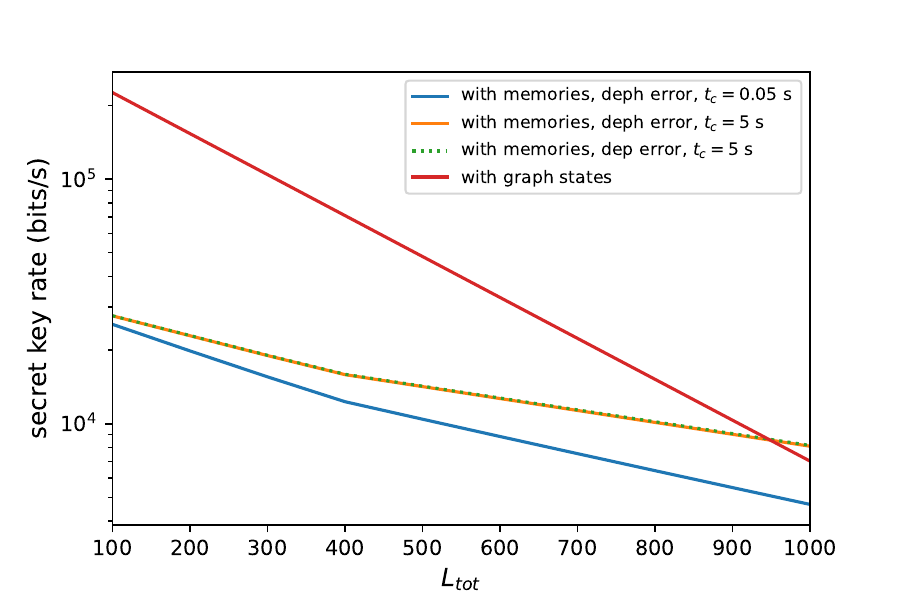}
    \caption{\label{figmcsc} The optimized SKRs of 3-loss cat codes for both approaches when the $4k$ photon losses are chosen as the desired syndrome measurement outcomes. The blue and orange solid curves show the SKRs of using quantum memories with dephasing channel, the red dotted curve shows SKR of using quantum memories with depolarizing channel, and the red sold curve shows the SKR of using graph states. The measurement error $1-p_m$ is fixed to be $0.1\%$.}
\end{figure}
We determine the elementary distances required to reach a cost coefficient $C^{\prime}=100$ versus the distance separating Alice and Bob for single-channel protocol with binomial codes (orange curve) and multi-channel protocol with cat codes (blue curve),
as shown in Figure \ref{figcpmc}. 
Figure \ref{figcpmc} shows that the elementary distance at $C^\prime=100$ is larger in the case of the multi-channel protocol compared to the case of the single-channel protocol embedded with binomial codes.
For instance, at a total distance $L_{tot}=1000$ km, the elementary distance to reach $C^{\prime}=100$ is 0.64 km for single-channel protocol with binomial codes while it can be extended to 1.25 km for multi-channel protocol with cat codes. 
Although the SKR of binomial codes is higher than cat codes, as shown in the inset of Figure \ref{figcpmc}, the number of resources needed are much higher in the protocol with binomial codes. 
This can be explained by considering that in the case of the single-channel protocol with binomial code, the number of stations is higher than the multi-channel case because the elementary distance is smaller. This will also enhance the number of auxiliary resources at the stations, such as detectors, cavities, and devices used to trap the atoms or ions.

To compare the performance of the multi-channel protocol with other existing quantum repeater protocols, we calculate the SKRs when the photonic modes are encoded with the 3-loss cat codes both by using quantum memories or graph states. As shown in Figure \ref{figmcsc}, with measurement error $1-p_m=0.1\%$, the SKR of 3-loss codes can reach $2\times 10^5$ bits/s with total distance $L_{tot}=100$ km and $7\times 10^3$ with $L_{tot}=1000$ km. 
We compare these results with the ones obtained in \cite{Borregaard2020}, in which a third generation quantum repeater protocol has been recently introduced. We see that the performance of the multi-channel protocol embedded with 3-loss cat codes can achieve slightly higher performance as long as the error probability $\epsilon_r$ in \cite{Borregaard2020} is equal to $0.05\%$.
In addition, the protocol proposed in \cite{Borregaard2020} requires hundreds of photons for encoding while the protocol proposed in this work only requires one single light mode. 

\section{\label{con}Concluding Discussion}

In this work we have presented a multi-channel QR protocol to
to select a favorite outcome of the syndrome measurements 
performed in each appropriate station. This method increases tremendously the SKR of the cat codes. 
For a fixed value of the SKR we estimate the trade-off between the coherence time, modeled as dephasing or 
depolarizing channel, and the gate efficiency. The coherence times required, if the depolarizing channel is considered, 
are near 1 minute, which are feasible within cavity-QED regime \cite{Wang2017}. If the dephasing channel is considered, 
the coherence times can be roughly 3 orders of magnitude smaller.
Furthermore, the SKR is comparable to the existing 3rd generation quantum repeater protocol but fewer optical resources are required. Thus, we expect the protocol proposed here based on RSBCs might have a more feasible implementation.
However, the local losses will accumulate at every elementary 
station and inevitably decrease the SKR. 
Therefore, further error-correcting 
elements such as additional quantum error correction codes on 
the atomic spin qubits need to be considered in future works. 

\begin{acknowledgments}
    We thank Thomas R. Scruby, Shin Nishio, Hon Wai Hana Lau and Nicholas Connolly for helpful discussions.
    This work was supported by the Moonshot R\&D Program Grants JPMJMS2061 \& JPMJMS226C, the JSPS KAKENHI Grant No. 21H04880 \& 24K07485 and JST, the establishment of university fellowships towards the creation of science technology innovation, Grant Number JPMJFS2136.
\end{acknowledgments}

\section*{Conflict of Interest}
The authors declare no conflict of interest.

\appendix
\section{\label{cvrs}The creation of graph states with nodes as RSBCs}

As mentioned in Section \ref{cs}, it is possible to create the graph state with nodes as RSBCs by employing the light-matter interaction based on cavity-QED.
A usual graph state is generated while each node is prepared in $\ket{+}$ state with CZ gates performed on the connections that one wants to make. However, due to the fact that the codewords of RSBCs defined in \cite{Li2023a} are not orthogonal and the light-matter interaction based on cavity-QED acts as a logical CNOT gate on these codes and the matter qubits instead of a CZ gate, the hybrid light-matter graph can not be generated as the usual way with the encoding and the light-matter interaction considered in this thesis. Here we will describe the equivalent method to create the graph states. Let us start with the simplest case where there are only four nodes with two of them being the matter qubits and the other two being the light modes encoded with RSBCs. The two matter qubits, B and C, are prepared in the superposition state $\ket{+}=(\ket{\uparrow}+\ket{\downarrow})/2$ while the light modes, A and D, are prepared in the logical 0 state $\ket{\bar{0}}$.

\begin{figure}[]
    \center
    \includegraphics[width=0.47\textwidth]{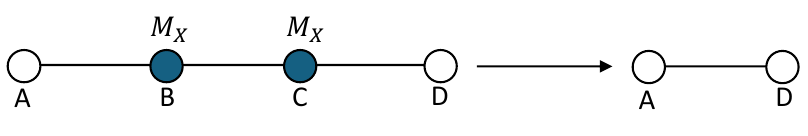}
    \caption{\label{figcslm} The measurement on $X$ basis performed on the matter qubits will remove them and connect their neighboring nodes as RSBCs. The blue circles are the trapped atoms/ions and the white circles are the light modes encoded with RSBCs. $M_X$s represent the measurements on $X$ basis.
    }
\end{figure}

As shown in Figure \ref{figcslm}, the light-matter interaction is applied on A and B as well as C and D. Then a CZ gate will be performed on B and C. The state after these operations can be written as
\begin{widetext}
    \begin{eqnarray}
        \label{eqa2.15}
        \begin{aligned}
        \ket{\psi}_{ABCD}=&{CZ}_{BC}(\ket{\bar{0}}_A\ket{\uparrow}_B+\ket{\bar{1}}_A\ket{\downarrow}_B)(\ket{\uparrow}_C\ket{\bar{0}}_D+\ket{\downarrow}_C\ket{\bar{1}}_D)/2\\
        =&(\ket{\bar{0}}_A\ket{\uparrow}_B\ket{\uparrow}_C\ket{\bar{0}}_D+\ket{\bar{1}}_A\ket{\downarrow}_B\ket{\uparrow}_C\ket{\bar{0}}_D+\ket{\bar{0}}_A\ket{\uparrow}_B\ket{\downarrow}_C\ket{\bar{1}}_D-\ket{\bar{1}}_A\ket{\downarrow}_B\ket{\downarrow}_C\ket{\bar{1}}_D)/2.
        \end{aligned}
    \end{eqnarray}
\end{widetext}
Then the measurements on X basis will be performed on both B and C nodes. Assuming that both the outcomes of the measurements yield $\ket{+}$ states. The state will become
\begin{eqnarray}
    \label{eqa2.16}
    \begin{aligned}
        \ket{\psi}_{AD}=&(\ket{\bar{0}}_A\ket{\bar{0}}_D+\ket{\bar{1}}_A\ket{\bar{0}}_D\\
        &+\ket{\bar{0}}_A\ket{\bar{1}}_D-\ket{\bar{1}}_A\ket{\bar{1}}_D)/2,
    \end{aligned}
\end{eqnarray}
which is analogous to the definition of graph state with two nodes.
For different outcomes of the measurements, an error correction will be performed on the remaining nodes to obtain the states needed. However, due to the lack of logical gates on the RSBCs, the information of the outcomes will be kept and sent to Alice and Bob. They will use this information to update the corresponding Pauli frame. This process can then be extended to graph states with more nodes. Thus, the graph states with nodes as RSBCs can be created with the specific light-matter interaction based on cavity-QED and local gates and measurements on the matter qubits, which are realizable in the experiment.
\section{\label{acs} Fidelity Analysis of the Multi-Channel Protocol Using Graph States}
Assuming there are two channels in use of the quantum repeater system. Then the graph states can be illustrated as shown in the right part of Figure \ref{figcs}, which
can be written as
\begin{equation}
    \label{eqa2.2}
    \begin{aligned}
        \ket{\phi_{graph}}_{1234}=&\frac{1}{N_c}(\ket{\overline{0000}}+\ket{\overline{0100}}+\ket{\overline{0010}}-\ket{\overline{0110}}\\
        &+\ket{\overline{0001}}-\ket{\overline{0011}}+\ket{\overline{0101}}+\ket{\overline{0111}}\\
        &+\ket{\overline{1000}}-\ket{\overline{1100}}+\ket{\overline{1010}}+\ket{\overline{1110}}\\
        &-\ket{\overline{1001}}+\ket{\overline{1101}}
        +\ket{\overline{1011}}+\ket{\overline{1111}}).
    \end{aligned}
\end{equation}
where $\ket{\bar{0}}$ and $\ket{\bar{1}}$ are the codewords of the cat codes with unorthogonal encoding, i.e., 
\begin{eqnarray}
    \label{eqa2.3}
    \ket{\bar{0}}=\frac{\ket{\alpha}+\ket{-\alpha}}{\sqrt{N_+}},\qquad
    \ket{\bar{1}}=\frac{\ket{i\alpha}+\ket{-i\alpha}}{\sqrt{N_+}}.
\end{eqnarray}
Here $\ket{\bar{+}}$ and $\ket{\bar{-}}$ are defined as $\ket{\bar{+}}=\ket{\bar{0}}+\ket{\bar{1}}$ and $\ket{\bar{-}}=\ket{\bar{0}}-\ket{\bar{1}}$. They are not normalized but $\ket{\phi_{cluster}}$ will be normalized in the end with the normalization constant $N_c$.  Then after the channel loss, the state will become
\begin{widetext}
    \begin{eqnarray}
        \label{eqa2.4}
        \begin{aligned}
            \hat{\rho}'=&\sum_{l=0}^{\infty}{\hat{A}_{4,l}(\sum_{k=0}^{\infty}{\hat{A}_{3,k}(\sum_{j=0}^{\infty}{\hat{A}_{2,j}(\sum_{j=0}^{\infty}{\hat{A}_{1,i}\ket{\phi_{cluster}}\bra{\phi_{cluster}}\hat{A}_{1,i}^\dagger})\hat{A}_{2,j}^\dagger})\hat{A}_{3,k}^\dagger})\hat{A}_{4,l}^\dagger}\\
            =&\sum_{i,j,k,l=0}^{\infty}{\hat{A}_{4,l}\hat{A}_{3,k}\hat{A}_{2,j}\hat{A}_{1,i}\ket\phi_{cluster}\bra{\phi_{cluster}}\hat{A}_{1,i}^\dagger\hat{A}_{2,j}^\dagger\hat{A}_{3,k}^\dagger\hat{A}_{4,l}^\dagger}
        \end{aligned}
    \end{eqnarray}
\end{widetext}
Here $\hat{A}_{i,k}$ is the Kraus operator of the amplitude damping channel for i-th mode. For any light mode, 
$\hat{A}_k=\sqrt{(1-\eta)^k/k!}\sqrt{\eta}^{\hat{n}}\hat{a}^k$. 
Now assuming that all the light modes are encoded in 1-loss cat code and after the syndrome measurement, the outcomes are even photon losses for mode 1 and 4, odd photon losses for mode 2 and 3. It means the state will be projected on
\begin{widetext}
    \begin{eqnarray}
        \label{eqa2.5}
        \hat{\rho}''=\frac{1}{N_c''}(\sum_{i,j,k,l=0}^{\infty}{\hat{A}_{4,2l}\hat{A}_{3,2k+1}\hat{A}_{2,2j+1}\hat{A}_{1,2i}\ket\phi_{cluster}\bra{\phi_{cluster}}\hat{A}_{1,2i}^\dagger\hat{A}_{2,2j+1}^\dagger\hat{A}_{3,2k+1}^\dagger\hat{A}_{4,2l}^\dagger}),\\
        where\ N_c''=tr{\sum_{i,j,k,l=0}^{\infty}{\hat{A}_{4,2l}\hat{A}_{3,2k+1}\hat{A}_{2,2j+1}\hat{A}_{1,2i}\ket\phi_{cluster}\bra{\phi_{cluster}}\hat{A}_{1,2i}^\dagger\hat{A}_{2,2j+1}^\dagger\hat{A}_{3,2k+1}^\dagger\hat{A}_{4,2l}^\dagger}}.
    \end{eqnarray}
\end{widetext}

For 1-loss cat code, we have the relation
\\
\\
\\
\\
\begin{eqnarray}
    \label{eqa2.6}
    \begin{aligned}
        \hat{A}_{2m}\ket{\bar{0}}=&\sqrt{\frac{\cosh(\eta\alpha^2)}{\cosh(\alpha^2)}}\frac{\sqrt{1-\eta}^{2m}\alpha^{2m}}{\sqrt{(2m)!}}\ket{\tilde{0}_+},\\
        \hat{A}_{2m}\ket{\bar{1}}=&\sqrt{\frac{\cosh(\eta\alpha^2)}{\cosh(\alpha^2)}}\frac{\sqrt{1-\eta}^{2m}\alpha^{2m}(-1)^m}{\sqrt{(2m)!}}\ket{\tilde{1}_+},\\
        \hat{A}_{2m+1}\ket{\bar{1}}=&\sqrt{\frac{\sinh(\eta\alpha^2)}{\cosh(\alpha^2)}}\frac{\sqrt{1-\eta}^{2m+1}\alpha^{2m+1}}{\sqrt{(2m+1)!}}\ket{\tilde{1}_+},\\
        \hat{A}_{2m+1}\ket{\bar{1}}=&\sqrt{\frac{\sinh(\eta\alpha^2)}{\cosh(\alpha^2)}}\frac{\sqrt{1-\eta}^{2m+1}\alpha^{2m+1}i^{2m+1}}{\sqrt{(2m+1)!}}\ket{\tilde{1}_+},\\
        &\ \ \ for\ any\ m\in\mathcal{N},
    \end{aligned}
\end{eqnarray}
where
\begin{eqnarray}
    \label{eqa2.7}
    \begin{aligned}
        \ket{\tilde{0}_+}=&\frac{\ket{\sqrt{\eta}\alpha}+\ket{-\sqrt{\eta}\alpha}}{\sqrt{N_+'}},\\
        \ket{\tilde{1}_+}=&\frac{\ket{i\sqrt{\eta}\alpha}+\ket{-i\sqrt{\eta}\alpha}}{\sqrt{N_+'}},\\
        \ket{\tilde{0}_-}=&\frac{\ket{\sqrt{\eta}\alpha}-\ket{-\sqrt{\eta}\alpha}}{\sqrt{N_-'}},\\
        \ket{\tilde{1}_-}=&\frac{\ket{i\sqrt{\eta}\alpha}-\ket{-i\sqrt{\eta}\alpha}}{\sqrt{N_-'}}.\\
    \end{aligned}
\end{eqnarray}
The normalization constants $\sqrt{N_+'}$ and $\sqrt{N_-'}$ can be written as
\begin{eqnarray}
    \label{eqa2.8}
    \begin{aligned}
        N_+'=&2+2\exp(-2\eta\alpha^2)=4\exp(-\eta\alpha^2)\cosh(\eta\alpha^2),\\
        N_-'=&2-2\exp(-2\eta\alpha^2)=4\exp(-\eta\alpha^2)\sinh(\eta\alpha^2).\\
    \end{aligned}
\end{eqnarray}
For simplicity, here we define 
\begin{eqnarray}
    \label{eqa2.9}
    \begin{aligned}
        C_m=&\sqrt{\frac{\cosh(\eta\alpha^2)}{\cosh(\alpha^2)}}\frac{\sqrt{1-\eta}^{2m}\alpha^{2m}}{\sqrt{(2m)!}},\\
        D_m=&\sqrt{\frac{\sinh(\eta\alpha^2)}{\cosh(\alpha^2)}}\frac{\sqrt{1-\eta}^{2m+1}\alpha^{2m+1}}{\sqrt{(2m)!}}.\\
    \end{aligned}
\end{eqnarray}
Then we have $\hat{A}_{2m}\ket{\bar{0}}=C_m\ket{\tilde{0}_+},\hat{A}_{2m}\ket{\bar{1}}=C_m(-1)^m\ket{\tilde{1}_+},\hat{A}_{2m+1}\ket{\bar{0}}=D_m\ket{\tilde{0}_-},\hat{A}_{2m+1}\ket{\bar{1}}=D_m i^{2m+1}\ket{\tilde{1}_+}$.

With all the relations above, now let us look at one of the terms in $\hat{\rho}''$,
\begin{eqnarray}
    \label{eqa2.10}
    \begin{aligned}
        &\hat{A}_{4,2l}\hat{A}_{3,2k+1}\hat{A}_{2,2j+1}\hat{A}_{1,2i}\ket{\overline{0+00}}\\
        =&C_lD_kD_jC_i\ket{\tilde{0}_+}(\ket{\tilde{0}_-}+i^{2k+1}\ket{\tilde{1}_-})\ket{\tilde{0}_-}\ket{\tilde{0}_+}.
    \end{aligned}
\end{eqnarray}
Then in order to get rid of the light modes with odd photon losses, a Pauli-Z measurement will be performed on each of them (on mode 2 and 3). Assuming the outcomes of the measurement are all +1 (the states will be projected to $\ket{\tilde{0}_-}$), then the resulting state after the measurement will be 
\begin{eqnarray}
    \label{eqa2.11}
    \hat{\rho}'''=\frac{1}{N_0'''}\sum_{i,j,k,l=0}^{\infty}{C_lD_kD_jC_i\ket{\tilde{\phi}_+}_{il}\bra{\tilde{\phi}_+}},
\end{eqnarray}
where
\begin{eqnarray}
    \label{eqa2.12}
    \begin{aligned}
        \ket{\tilde{\phi}_+}_{il}=&\ket{\tilde{0}_+\tilde{0}_+}+(-1)^l\ket{\tilde{0}_+\tilde{1}_+}
        +(-1)^i\ket{\tilde{1}_+\tilde{0}_+}\\
        &-(-1)^{i+l}\ket{\tilde{1}_+\tilde{1}_+},
    \end{aligned}
\end{eqnarray}
which is independent of the coefficients $D_kD_j$. (Here $\ket{\tilde{\phi}_+}_{il}$ is not normalized and $\rho'''$ will be normalized by the normaliza=on constant $N'''$.) Then we can rewrite $\rho'''$ as 
\begin{eqnarray}
    \label{eqa2.13}
    \hat{\rho}'''=\frac{1}{N_0'''}\sum_{i,l=0}^{\infty}{C_lC_i\ket{\tilde{\phi}_+}_{il}\bra{\tilde{\phi}_+}}\sum_{j,k=0}^{\infty}{D_kD_j}.
\end{eqnarray}
Then it can be seen that $\sum_{j,k=0}^{\infty}{D_kD_j}$ is just a global phase (norm), which can be subtracted when we normalize $\hat{\rho}'''$. So we can rewrite $\hat{\rho}'''$ without these terms as
\begin{eqnarray}
    \label{eqa2.14}
    \hat{\rho}'''=\frac{1}{N_0'''}\sum_{i,l=0}^{\infty}{C_lC_i\ket{\tilde{\phi}_+}_{il}\bra{\tilde{\phi}_+}}.
\end{eqnarray}
where $N'''=N_0'''/\sum_{j,k=0}^{\infty}{D_kD_j}$. Now we can see that $\hat{\rho}'''$ is equivalent to that we transmit the state $\ket{\overline{00}}_{14}+\ket{\overline{01}}_{14}+\ket{\overline{10}}_{14}-\ket{\overline{11}}_{14}$ through the channel and the outcomes of the syndrome measurements are even photon losses for both mode 1 and mode 4.

According to Equation \ref{eqa2.12}, $\ket{\tilde{\phi}_+}_{il}$ depends on the number parity of $i$ and $l$. So there are 4 cases now: 
both $i$ and $l$ are even; both of $i$ and $l$ are odd; $i$ is odd and $l$ is even; $i$ is even and $l$ is odd. Then 
we can rewrite $\hat{\rho}'''$ as
\begin{eqnarray}
    \label{eqa2.17}
    \begin{aligned}
        \hat{\rho}'''=&\frac{1}{N_0'''}(\sum_{i',l'=0}^{\infty}{C_{2l'}C_{2i'}\ket{\tilde{\phi}_+}_{2i',2l'}\bra{\tilde{\phi}_+}}\\
        &+\sum_{i',l'=0}^{\infty}{C_{2l'}C_{2i'+1}\ket{\tilde{\phi}_+}_{2i'+1,2l'}\bra{\tilde{\phi}_+}}\\
        &+\sum_{i',l'=0}^{\infty}{C_{2l'+1}C_{2i'}\ket{\tilde{\phi}_+}_{2i',2l'+1}\bra{\tilde{\phi}_+}}\\
        &+\sum_{i',l'=0}^{\infty}{C_{2l'+1}C_{2i'+1}\ket{\tilde{\phi}_+}_{2i'+1,2l'+1}\bra{\tilde{\phi}_+}}),
    \end{aligned}
\end{eqnarray}
where
\begin{eqnarray}
    \label{eqa2.18}
    \begin{aligned}
        \ket{\tilde{\phi}_+}_{2i',2l'}=&\ket{\tilde{0}_+\tilde{0}_+}+\ket{\tilde{0}_+\tilde{1}_+}
        +\ket{\tilde{1}_+\tilde{0}_+}\\
        &-\ket{\tilde{1}_+\tilde{1}_+},\\
        \ket{\tilde{\phi}_+}_{2i'+1,2l'}=&\ket{\tilde{0}_+\tilde{0}_+}+\ket{\tilde{0}_+\tilde{1}_+}-\ket{\tilde{1}_+\tilde{0}_+}\\
        &+\ket{\tilde{1}_+\tilde{1}_+},\\
        \ket{\tilde{\phi}_+}_{2i',2l'+1}=&\ket{\tilde{0}_+\tilde{0}_+}-\ket{\tilde{0}_+\tilde{1}_+}+\ket{\tilde{1}_+\tilde{0}_+}\\
        &+\ket{\tilde{1}_+\tilde{1}_+},\\
        \ket{\tilde{\phi}_+}_{2i'+1,2l'+1}=&\ket{\tilde{0}_+\tilde{0}_+}-\ket{\tilde{0}_+\tilde{1}_+}-\ket{\tilde{1}_+\tilde{0}_+}\\
        &-\ket{\tilde{1}_+\tilde{1}_+}.
    \end{aligned}
\end{eqnarray}
Here all the states in Eqation \ref{eqa2.18} are independent of the value of $i'$ and $l'$, so we can define
\begin{eqnarray}
    \label{eqa2.19}
    \begin{aligned}
        &\ket{\tilde{\phi}_+}_{2i',2l'}=\ket{\tilde{\phi}_+}_1,\qquad &&\ket{\tilde{\phi}_+}_{2i'+1,2l'}=\ket{\tilde{\phi}_+}_2,\\
        &\ket{\tilde{\phi}_+}_{2i',2l'+1}=\ket{\tilde{\phi}_+}_3, &&\ket{\tilde{\phi}_+}_{2i'+1,2l'+1}=\ket{\tilde{\phi}_+}_4.
    \end{aligned}    
\end{eqnarray}
Then we have 
\begin{eqnarray}
    \label{eqa2.20}
    \begin{aligned}
        \hat{\rho}'''=&\frac{1}{N_0'''}(\ket{\tilde{\phi}_+}_1\bra{\tilde{\phi}_+}\sum_{i',l'=0}^{\infty}{C_{2l'}C_{2i'}}\\
        &+\ket{\tilde{\phi}_+}_2\bra{\tilde{\phi}_+}\sum_{i',l'=0}^{\infty}{C_{2l'}C_{2i'+1}}\\
        &+\ket{\tilde{\phi}_+}_3\bra{\tilde{\phi}_+}\sum_{i',l'=0}^{\infty}{C_{2l'+1}C_{2i'}}\\
        &+\ket{\tilde{\phi}_+}_4\bra{\tilde{\phi}_+}\sum_{i',l'=0}^{\infty}{C_{2l'+1}C_{2i'+1}}),
    \end{aligned}
\end{eqnarray}
which is the mixture of 4 different pure states $\ket{\tilde{\phi}_+}_1$, $\ket{\tilde{\phi}_+}_2$, $\ket{\tilde{\phi}_+}_3$ and $\ket{\tilde{\phi}_+}_4$. Next, let us look at the resulting states after entanglement creation process for these 4 pure states according to 
the corresponding light-matter interaction. First, for $\ket{\tilde{\phi}_+}_1$, we have 
\begin{eqnarray}
    \label{eqa2.21}
    \begin{aligned}
        &\frac{1}{\sqrt{2}}(\ket{\uparrow}+\ket{\downarrow})\ket{\tilde{\phi}_+}_1\frac{1}{\sqrt{2}}(\ket{\uparrow}+\ket{\downarrow})\rightarrow\\
        [&\ket{\uparrow}(\ket{\tilde{0}_+\tilde{0}_+}+\ket{\tilde{0}_+\tilde{1}_+}+\ket{\tilde{1}_+\tilde{0}_+}-\ket{\tilde{1}_+\tilde{1}_+})\ket{\uparrow}\\
        +&\ket{\uparrow}(\ket{\tilde{0}_+\tilde{0}_+}+\ket{\tilde{0}_+\tilde{1}_+}-\ket{\tilde{1}_+\tilde{0}_+}+\ket{\tilde{1}_+\tilde{1}_+})\ket{\downarrow}\\
        +&\ket{\downarrow}(\ket{\tilde{0}_+\tilde{0}_+}-\ket{\tilde{0}_+\tilde{1}_+}+\ket{\tilde{1}_+\tilde{0}_+}+\ket{\tilde{1}_+\tilde{1}_+})\ket{\uparrow}\\
        +&\ket{\downarrow}(-\ket{\tilde{0}_+\tilde{0}_+}+\ket{\tilde{0}_+\tilde{1}_+}+\ket{\tilde{1}_+\tilde{0}_+}+\ket{\tilde{1}_+\tilde{1}_+})\ket{\downarrow}]/2.
    \end{aligned}
\end{eqnarray}
Then a state discrimination (equivalent to logical Z measurement here) on the light modes will be performed. Assuming the outcome is $\ket{\tilde{0}_+\tilde{0}_+}$, then the final atomic state is 

\begin{eqnarray}
    \label{eqa2.22}
    \begin{aligned}
        &\frac{\ket{\uparrow\uparrow}+\ket{\uparrow\downarrow}+\ket{\downarrow\uparrow}-\ket{\downarrow\downarrow}}{2}\\
        =&\frac{\ket{\uparrow+}+\ket{\downarrow-}}{2}\\
        =&\frac{\ket{+\uparrow}+\ket{-\downarrow}}{2},
    \end{aligned}
\end{eqnarray}
where $\ket{+}=(\ket{\uparrow}+\ket{\downarrow})/\sqrt{2},\ \ket{-}=(\ket{\uparrow}-\ket{\downarrow})/\sqrt{2}.$ If necessary, we can get one of the Bell states  $(\ket{\uparrow\uparrow}+\ket{\downarrow\downarrow})/\sqrt{2}$ by performing a Hadamard operation on one of the matter qubits. If the outcome is not $\ket{\tilde{0}_+\tilde{0}_+}$, the final atomic state will be the state in Equation \ref{eqa2.22} with Pauli-Z operation(s) on corresponding matter qubit(s). Similarly, for $\ket{\tilde{\phi}_+}_2$, $\ket{\tilde{\phi}_+}_3$ and $\ket{\tilde{\phi}_+}_4$, we have 
\begin{widetext}
    \begin{eqnarray}
        \label{eqa2.23}
        \begin{aligned}
            &\frac{1}{\sqrt{2}}(\ket{\uparrow}+\ket{\downarrow})\ket{\tilde{\phi}_+}_2\frac{1}{\sqrt{2}}(\ket{\uparrow}+\ket{\downarrow})\rightarrow\\
            [\ket{\uparrow}(\ket{\tilde{0}_+\tilde{0}_+}+\ket{\tilde{0}_+\tilde{1}_+}&-\ket{\tilde{1}_+\tilde{0}_+}+\ket{\tilde{1}_+\tilde{1}_+})\ket{\uparrow}+\ket{\uparrow}(\ket{\tilde{0}_+\tilde{0}_+}+\ket{\tilde{0}_+\tilde{1}_+}+\ket{\tilde{1}_+\tilde{0}_+}-\ket{\tilde{1}_+\tilde{1}_+})\ket{\downarrow}\\
            +\ket{\downarrow}(-\ket{\tilde{0}_+\tilde{0}_+}+\ket{\tilde{0}_+\tilde{1}_+}&+\ket{\tilde{1}_+\tilde{0}_+}+\ket{\tilde{1}_+\tilde{1}_+})\ket{\uparrow}+\ket{\downarrow}(\ket{\tilde{0}_+\tilde{0}_+}-\ket{\tilde{0}_+\tilde{1}_+}+\ket{\tilde{1}_+\tilde{0}_+}+\ket{\tilde{1}_+\tilde{1}_+})\ket{\downarrow}]/2,\\
            &\frac{1}{\sqrt{2}}(\ket{\uparrow}+\ket{\downarrow})\ket{\tilde{\phi}_+}_3\frac{1}{\sqrt{2}}(\ket{\uparrow}+\ket{\downarrow})\rightarrow\\
            [\ket{\uparrow}(\ket{\tilde{0}_+\tilde{0}_+}-\ket{\tilde{0}_+\tilde{1}_+}&+\ket{\tilde{1}_+\tilde{0}_+}+\ket{\tilde{1}_+\tilde{1}_+})\ket{\uparrow}+\ket{\uparrow}(-\ket{\tilde{0}_+\tilde{0}_+}+\ket{\tilde{0}_+\tilde{1}_+}+\ket{\tilde{1}_+\tilde{0}_+}+\ket{\tilde{1}_+\tilde{1}_+})\ket{\downarrow}\\
            +\ket{\downarrow}(\ket{\tilde{0}_+\tilde{0}_+}+\ket{\tilde{0}_+\tilde{1}_+}&+\ket{\tilde{1}_+\tilde{0}_+}-\ket{\tilde{1}_+\tilde{1}_+})\ket{\uparrow}+\ket{\downarrow}(\ket{\tilde{0}_+\tilde{0}_+}+\ket{\tilde{0}_+\tilde{1}_+}-\ket{\tilde{1}_+\tilde{0}_+}+\ket{\tilde{1}_+\tilde{1}_+})\ket{\downarrow}]/2,\\
            &\frac{1}{\sqrt{2}}(\ket{\uparrow}+\ket{\downarrow})\ket{\tilde{\phi}_+}_4\frac{1}{\sqrt{2}}(\ket{\uparrow}+\ket{\downarrow})\rightarrow\\
            [\ket{\uparrow}(\ket{\tilde{0}_+\tilde{0}_+}-\ket{\tilde{0}_+\tilde{1}_+}&-\ket{\tilde{1}_+\tilde{0}_+}-\ket{\tilde{1}_+\tilde{1}_+})\ket{\uparrow}+\ket{\uparrow}(-\ket{\tilde{0}_+\tilde{0}_+}+\ket{\tilde{0}_+\tilde{1}_+}-\ket{\tilde{1}_+\tilde{0}_+}-\ket{\tilde{1}_+\tilde{1}_+})\ket{\downarrow}\\
            +\ket{\downarrow}(-\ket{\tilde{0}_+\tilde{0}_+}-\ket{\tilde{0}_+\tilde{1}_+}&-\ket{\tilde{1}_+\tilde{0}_+}+\ket{\tilde{1}_+\tilde{1}_+})\ket{\uparrow}+\ket{\downarrow}(-\ket{\tilde{0}_+\tilde{0}_+}-\ket{\tilde{0}_+\tilde{1}_+}-\ket{\tilde{1}_+\tilde{0}_+}+\ket{\tilde{1}_+\tilde{1}_+})\ket{\downarrow}]/2.
        \end{aligned}
    \end{eqnarray}
\end{widetext}
So after the entanglement creation process (assuming the output of the state discrimination is $\ket{\tilde{0}_+\tilde{0}_+}$), the density matrix of the final atomic state is 
\begin{eqnarray}
    \label{eqa2.24}
    \begin{aligned}
        \hat{\rho}''''=&\frac{1}{N''''}[\frac{(\ket{\uparrow+}+\ket{\downarrow-})(\bra{\uparrow+}+\bra{\downarrow-})}{2}\sum_{i',l'=0}^{\infty}{C_{2l'}C_{2i'}}\\
        &+\frac{(\ket{\uparrow+}-\ket{\downarrow-})(\bra{\uparrow+}-\bra{\downarrow-})}{2}\sum_{i',l'=0}^{\infty}{C_{2l'}C_{2i'+1}}\\
        &+\frac{(\ket{\uparrow-}+\ket{\downarrow+})(\bra{\uparrow-}+\bra{\downarrow+})}{2}\sum_{i',l'=0}^{\infty}{C_{2l'+1}C_{2i'}}\\
        &+\frac{(\ket{\uparrow-}-\ket{\downarrow+})(\bra{\uparrow-}-\bra{\downarrow+})}{2}\sum_{i',l'=0}^{\infty}{C_{2l'+1}C_{2i'+1}}].
    \end{aligned}
\end{eqnarray}
Then if we set our target state as $\ket{\phi_t}=(\ket{\uparrow+}+\ket{\downarrow-})/\sqrt{2}$ (no-loss case), then the fidelity is 
\begin{eqnarray}
    \label{eqa2.25}
    F=\braket{\phi_t|\hat{\rho}''''|\phi_t}=\frac{1}{N''''}\sum_{i',l'=0}^{\infty}{C_{2l'}C_{2i'}},
\end{eqnarray}
where 

\begin{eqnarray}
    \label{eqa2.26}
    \begin{aligned}
        N''''=&\sum_{i',l'=0}^{\infty}{C_{2l'}C_{2i'}}
        +\sum_{i',l'=0}^{\infty}{C_{2l'}C_{2i'+1}}\\
        &+\sum_{i',l'=0}^{\infty}{C_{2l'+1}C_{2i'}}+\sum_{i',l'=0}^{\infty}{C_{2l'+1}C_{2i'+1}}\\
        =&\sum_{i,l=0}^{\infty}{C_{l}C_{i}}.
    \end{aligned}
\end{eqnarray}
This fidelity is exactly the same as if we transmit the state $\ket{\overline{00}}_{14}+\ket{\overline{11}}_{14}$ through the channel.

Next, we will show how the state evolves in a syndrome measurement process.
According to Equation \ref{eqa2.4}, if there is no syndrome measurement, then the state is
\begin{widetext}
    \begin{eqnarray}
        \label{eqa2.27}
        \hat{\rho}'=\sum_{i,j,k,l=0}^{\infty}{\hat{A}_{4,l}\hat{A}_{3,k}\hat{A}_{2,j}\hat{A}_{1,i}\ket\phi_{cluster}\bra{\phi}\hat{A}_{1,i}^\dagger\hat{A}_{2,j}^\dagger\hat{A}_{3,k}^\dagger\hat{A}_{4,l}^\dagger},
    \end{eqnarray}
\end{widetext}  
so $i,j,k,l$ here can be either even or odd. Then after the logical Z measurement on mode 2 and 3 and assuming the outcomes of the measurement are all +1 (the states will be projected to $\ket{\tilde{0}_+\tilde{0}_+}$ or $\ket{\tilde{0}_-\tilde{0}_-}$), the state will become 
\begin{widetext}
    \begin{eqnarray}
        \label{eqa2.28}
        \begin{aligned}
            \hat{\rho}_{cl}'=&\sum_{i,j,k,l=0}^{\infty}{C_lD_kD_jC_i\ket{\tilde{\phi}_+}_{il}\bra{\tilde{\phi}_+}+C_lC_kD_jC_i\ket{\tilde{\phi}_+}_{il}\bra{\tilde{\phi}_+}}+C_lD_kC_jC_i\ket{\tilde{\phi}_+}_{il}\bra{\tilde{\phi}_+}+C_lC_kC_jC_i\ket{\tilde{\phi}_+}_{il}\bra{\tilde{\phi}_+}+...\\
            =&\sum_{i,l=0}^{\infty}{C_lC_i\ket{\tilde{\phi}_+}_{il}\bra{\tilde{\phi}_+}}\sum_{j,k=0}^{\infty}{(D_kD_j+C_kD_j+D_kC_j+C_kC_j)}+...
        \end{aligned}
    \end{eqnarray}
\end{widetext}
Here only the  even loss terms for mode 1 and 4 were shown, but similarly, the odd loss term (both mode 1 and 4 lose odd number of photons) is 
\begin{widetext}
    \begin{eqnarray}
        \label{eqa2.29}
        \begin{aligned}
            \sum_{i,l=0}^{\infty}{D_lD_i\ket{\tilde{\phi}_+}_{il}\bra{\tilde{\phi}_+}}\sum_{j,k=0}^{\infty}{(D_kD_j+C_kD_j+D_kC_j+C_kC_j)},
        \end{aligned}
    \end{eqnarray}
where
    \begin{eqnarray}
        \label{eqa2.30}
        \begin{aligned}
        \sum_{j,k=0}^{\infty}{(D_kD_j+C_kD_j+D_kC_j+C_kC_j)}=&\sum_{j,k=0}^{\infty}{(D_k+C_k)(D_j+C_j)}\\
        =&\sum_{k=0}^{\infty}{(D_k+C_k)}\sum_{j=0}^{\infty}{(D_j+C_j)}=1,
        \end{aligned}
    \end{eqnarray}
\end{widetext}
Thus, $\hat{\rho}_{cl}'$ here is equivalent to the final density that one would obtain in the case that the state $\ket{\overline{00}}_{14}+\ket{\overline{01}}_{14}+\ket{\overline{10}}_{14}-\ket{\overline{11}}_{14}$ through the channel.

\section{\label{cvf}QBERs of the Multi-Channel Protocol Using Quantum Memories}
In this subsection, we will show the derivation of the fidelities for the multi-channel protocol using QMs. As mentioned in the main text, the depolarizing and dephasing channels are considered when the QMs are used. 
When a quantum state with density matrix $\hat{\rho}$ goes through a depolarizing channel, the density matrix will evolve as below \cite{Nielsen2010}:
\begin{eqnarray}
    \label{eqac1}
    \hat{\rho}'_{dp}=p_{dp}\hat{I}/d+(1-p_{dp})\hat{\rho},
\end{eqnarray}
where $\hat{I}/d$ is the completely mixed state with $d$ being the dimension of the Hilbert space of the system and $p_{dp}$ is the probability that the state is depolarized. 
For a dephasing channel, the evolution of the quantum state can be written as
\begin{eqnarray}
    \label{eqac2}
    \hat{\rho}'_{dph}=p_{dph}\hat{Z}\hat{\rho}\hat{Z}+(1-p_{dph})\hat{\rho},
\end{eqnarray}
where $\hat{Z}$ is the Pauli-Z operator and $p_{dph}$ is the probability that a phase-flip error occurs. 
The probability $p_{dp}$ and $p_{dph}$ are related to the coherence time of a quantum memory, which can be written as
\begin{eqnarray}
    \label{eqac3}
    p_{dp/dph}=e^{-t/t_c},
\end{eqnarray}
where $t_c$ is the coherence time of the QM and $t$ is the evolution time of the quantum state.
In the multi-channel protocol using QMs, the depolarizing or dephasing errors occur after the light modes are sent from $ES_1$s and before the entanglement swapping at $ES_1$s. Thus, the duration of the quantum state through such a depolarizing or dephasing channel is given by $t_w=2L_0/c$.
In the case that the syndrome measurement outcome is even photon losses when the information is encoded using 1-loss cat codes, the density operator of the atomic state at each elementary link after the entanglement creation process can be written as \cite{Li2023a}
\begin{eqnarray}
    \label{eqac4}
    \hat{\rho}_0=F_0\ket{\Phi^+}\bra{\Phi^+}+(1-F_0)\ket{\Phi^-}\bra{\Phi^-},
\end{eqnarray}
where $\ket{\Phi^\pm}=(\ket{\uparrow\uparrow}\pm\ket{\downarrow\downarrow})/\sqrt{2}$ are the atomic Bell states and $F_0$ is the fidelity of the elementary link (More details about $F_0$ can be found in~\cite{Li2023a}). Then if the state is subject to a depolarizing channel, it will become
\begin{eqnarray}
    \label{eqac5}
    \hat{\rho}'_{0,dp}=p_{dp}\hat{I}/4+(1-p_{dp})\hat{\rho}_0,
\end{eqnarray}
Next, entanglement swapping will be performed at each elementary stations. To obtain the resulting state after the entanglement swapping, one can think of the scenario that one of the initial state before the entanglement swapping is the maximally mixed state $\hat{I}/2$. No matter what the other state is, the resulting state will be $\hat{I}/2$ after the entanglement swapping operation. Thus, the final state after the entanglement swapping performed at each elementary station will become
\begin{eqnarray}
    \label{eqac6}
    \hat{\rho}'_{f,dp}=[1-(1-p_{dp})^n]\hat{I}/4+(1-p_{dp})^n\hat{\rho}_f,
\end{eqnarray}
where $\hat{\rho}_f$ is the resulting state after the entanglement swapping assuming the initial state at each elementary station is $\hat{\rho}_0$. It can be shown that $\hat{\rho}_f$ is given by
\begin{eqnarray}
    \label{eqac7}
    \begin{aligned}
        \hat{\rho}_f=&\left[\frac{1+(2F_0-1)^n}{2}\right]\ket{\Phi^+}\bra{\Phi^+}\\
        &+\left[\frac{1-(2F_0-1)^n}{2}\right]\ket{\Phi^-}\bra{\Phi^-}.
    \end{aligned}
\end{eqnarray}
Then the QBERs can be written as
\begin{eqnarray}
    \label{eqac8}
    \begin{aligned}
        e_x^{dp}=&\frac{\left[1-(2F_0-1)^n\right](1-p_{dp})^n+1-(1-p_{dp})^n}{2},\\
        &\quad \quad \quad \quad e_z^{dp}=\frac{1-(1-p_{dp})^n}{2}.
    \end{aligned}
\end{eqnarray}
If the state is subject to a dephasing channel, it will become
\begin{eqnarray}
    \label{eqac9}
    \hat{\rho}'_{0,dph}=p_{dph}\hat{Z}_2\hat{\rho}_0\hat{Z}_2+(1-p_{dph})\hat{\rho}_0,
\end{eqnarray}
where $\hat{Z}_2$ is the $Z$ operator in a two-qubit system. Since only one of the qubits (the one at $ES_1$) will be stored in a QM and goes through the dephasing channel, we have $\hat{Z}_2=\hat{Z}_1\otimes\hat{I}$ with $\hat{Z}_1$ being the Z operator for a single qubit. Thus, we have $\hat{Z}_2\ket{\Phi_\pm}=\ket{\Phi_\mp}$ and $\hat{Z}_2\hat{Z}_2=\hat{I}$. Then if even number of the states in the elementary stations are subject to the operation of  $\hat{Z}_2$, after the entanglement swapping, the operation of $\hat{Z}_2$ will be cancelled out. In the other hand, if odd number of the states in the elementary stations are subject to the operation of  $\hat{Z}_2$, then the final state is subject to an operation of $\hat{Z}_2$.
Thus, the final state is given by
\begin{eqnarray}
    \label{eqac10}
    \begin{aligned}
        \hat{\rho}'_{f,dph}=&\left[\frac{1-(1-2p_{dph})^n}{2}\right]\hat{Z}_2\hat{\rho}_f\hat{Z}_2\\
        &+\left[\frac{1+(1-2p_{dph})^n}{2}\right]\hat{\rho}_f.
    \end{aligned}
\end{eqnarray}
Then according to Equation \ref{eqac7}, the QBERs can be written as
\begin{eqnarray}
    \label{eqac11}
    \begin{aligned}
        e_x^{dph}=&\left[\frac{1-(1-2p_{dph})^n}{2}\right]\left[\frac{1+(2F_0-1)^n}{2}\right]\\
        &+\left[\frac{1+(1-2p_{dph})^n}{2}\right]\left[\frac{1-(2F_0-1)^n}{2}\right],\\
        &\quad \quad \quad \quad \quad e_z^{dph}=0.
    \end{aligned}
\end{eqnarray}

\bibliography{QP2}
\end{document}